\documentclass[pra,letterpaper,preprint,showpacs,superscriptaddress,floatfix]{revtex4}
\usepackage{graphicx,psfrag,amsmath,amssymb,amsfonts,bbm,latexsym,color,dcolumn,bm}

\begin{document}
\title{Signatures of macroscopic quantum coherence in ultracold dilute
  Fermi gases}

\author{Roberto Onofrio}
\affiliation{Department of Physics and Astronomy, 
Dartmouth College, 6127 Wilder Laboratory, Hanover, NH 03755}
\affiliation{Dipartimento di Fisica ``G. Galilei'',
 Universit\`a di Padova, Via Marzolo 8, Padova 35131, Italy}
\affiliation{Center for Statistical Mechanics and Complexity, INFM, Unit\`a di Roma 1, Roma 00185, Italy}

\author{Carlo Presilla}
\affiliation{Dipartimento di Fisica, Universit\`a di Roma ``La Sapienza'',
Piazzale A. Moro 2, Roma 00185, Italy}
\affiliation{Center for Statistical Mechanics and Complexity, INFM, Unit\`a di Roma 1, Roma 00185, Italy}
\affiliation{INFN, Sezione di Roma 1, Roma 00185, Italy}


\begin{abstract}
We propose a double-well configuration for optical trapping 
of ultracold two-species Fermi-Bose atomic mixtures. 
Two signatures of macroscopic quantum coherence attributable to a 
superfluid phase transition for the Fermi gas are analyzed. 
The first signature is based upon tunneling of Fermi pairs when the 
power of the deconfining laser beam is significantly reduced. 
The second relies on the observation of interference 
fringes in a regime where the fermions are trapped in two 
sharply separated minima of the potential.
Both signatures rely on small decoherence times for the Fermi 
samples, which should be possible by reaching low temperatures 
using a Bose gas as a refrigerator, and a bichromatic optical dipole 
trap for confinement, with optimal heat capacity matching 
between the two species.
\end{abstract}

\maketitle

\section{Introduction}
Degenerate Fermi gases are rather ubiquitous in nature at both the 
microscopic and macroscopic level, from nuclear matter to neutron stars. 
Studies of both their non-interacting and interacting features 
allow for the understanding of a wealth of physical phenomena 
occurring in the mesoscopic realm, in particular superconductivity. 
More recently, the possibility to cool  dilute samples of Fermi gases 
below the microkelvin range has opened up a novel route to identify
some of the fundamental features underlying interacting 
many-body Fermi systems \cite{Stoof}. 
While Pauli blocking \cite{DeMarco} and Fermi pressure \cite{Truscott,Schreck} 
have been already evidenced, focus on interacting properties has
recently led to interesting effects in Fermi-Bose mixtures
\cite{Inguscio}, and in two-component Fermi gases \cite{Ohara}. 
In particular, in the latter case evidence has been reported for 
anisotropic free expansions of a Fermi cloud when this is brought 
to a deep degenerate regime. 
The data have been interpreted in terms of a superfluid 
state of the Fermi gas, as predicted in \cite{Menotti}. 
However, alternative interpretations in terms of hydrodynamic 
behaviour of a high density Fermi gas are also plausible \cite{Gupta}.
More recently, various groups have used resonant superfluidity 
\cite{Holland,Timmermans,Chiofalo,Ohashi} to explore the BEC-BCS 
crossover \cite{Regal,Ketterle,Grimm}, with various claims for 
the formation of bound states of many body character, as expected 
for instance by BCS-like couplings, based on the dynamics of 
formation of Fermi pairs and on collective properties \cite{Thomas}.  
This reminds of the previous situation of degenerate Bose gases, when 
various indirect evidences were collected for the existence of a 
macroscopic quantum state by studying collective properties. 
The final evidence was only achieved by explicitely showing quantum 
coherence \cite{Andrews} and, thereafter, macroscopic quantum
transport phenomena like superfluidity \cite{Raman} and 
quantized vortices \cite{Matthews,Fetter}.
Analogously, we do expect the coherence of the macroscopic wave
function associated to a Cooper-paired state of Fermi atoms to play 
an important role to assess its superfluid nature. 
In this paper, we discuss a configuration for an optical dipole trap 
that could allow for quantitative studies of quantum coherence in 
an ultracold Fermi gas. In Section II we describe a geometry 
for an optical dipole trap which creates a bistable potential for 
both the Fermi species and the Bose species necessary to
sympathetically cool the Fermi gas. In Section III 
we discuss possible signatures for macroscopic quantum coherence 
through tunneling phenomena in a regime where the laser intensity of the blue-detuned beam 
is kept low.  In Section IV we describe interference experiments which 
should be able to disentangle between a BCS or a BEC regime 
for the degenerate Fermi gas  by observing the
dynamics of the fringe visibility during the free expansion of the 
clouds. Macroscopic coherence in itself does not rely on the Fermi gas being in an 
effective BEC or a BCS state, as correlated Fermi pairs, 
either in a {\sl molecular} state (BEC limit) or a {\sl many-body} 
state (BCS limit) always behave as quantum coherent systems \cite{McDonald}. 
However, for the interference fringe experiment and BCS-paired
fermions, a sudden loss of fringe visibility is expected 
for large times of flight, while such a loss is not expected 
in the case of fermions coupled in a molecular state.  
Potential decoherence sources and some technical difficulties 
to be overcome are then discussed in the conclusive Section V.

\section{Double-well bichromatic optical dipole traps}
The configuration we analyze relies on using an optical dipole
trap made of focused red-detuned beams for trapping the atoms, and  
further blue-detuned beams for their selective deconfinement. 
Such a bichromatic optical dipole trap could allow to achieve a deep 
degenerate regime for a Fermi gas when the latter is sympathetically
cooled through a Bose gas undergoing evaporative cooling \cite{Onofrio}. 

Let us consider an optical dipole trap consisting of a single
red-detuned beam (optical source 1) propagating along the $x$-axis, 
and a single blue-detuned beam (optical source 2), 
also focused on the same spot, propagating along
the orthogonal axis $y$. 
The resulting effective potential experienced by the atoms of species 
$\alpha$ ($\alpha=\mathrm{f}$ for fermions and $\alpha=\mathrm{b}$ for bosons)
can be written as:
\begin{eqnarray}
U_\alpha(x,y,z)&=&-\frac{\hbar \Gamma_\alpha^2}
{4 \pi I_\alpha^{\mathrm sat}} 
\left[ 
\frac{T_1^\alpha P_1}{w_1^2} ~
\frac{\exp \left( \frac{-2\left(y^2+z^2\right)}
{w_1^2\left(1+x^2/R_1^2\right)}\right)}{1+x^2/R_1^2}
\right. \nonumber \\ && \qquad + \left.
\frac{T_2^\alpha P_2}{w_2^2} ~
\frac{\exp \left( \frac{-2\left(x^2+z^2\right)}
{w_2^2\left(1+y^2/R_2^2\right)}\right)}{1+y^2/R_2^2} 
\right] ,
\label{potential}
\end{eqnarray}
where $T_i^\alpha=1/(\Omega_\alpha-\Omega_i)+1/(\Omega_\alpha+\Omega_i)$ 
is a parameter related to the detuning between the atomic 
transition angular frequencies $\Omega_\alpha=2 \pi c/\lambda_\alpha$
and the laser beam angular frequencies $\Omega_i=2 \pi c/\lambda_i$ 
($\lambda_\alpha$ and $\lambda_i$ being the atomic transition and 
laser beam wavelengths, respectively), $P_i$ and $w_i$ are 
power and waist of the laser beams, $R_i=\pi w_i^2/\lambda_i$ 
their Rayleigh ranges, and $I_\alpha^{\mathrm sat}=\hbar \Omega_\alpha^3 
\Gamma_\alpha/12 \pi c^2$ is the saturation intensity for the atomic transition.
The potential (\ref{potential}) is well approximated 
by a second order Taylor expansion around the $x$ axis  
\begin{equation}
U_\alpha(x,y,z) \simeq U_\alpha(\bm{x})+
\frac{1}{2} 
\partial^2_y U_\alpha(\bm{x}) y^2+
\frac{1}{2} 
\partial^2_z U_\alpha(\bm{x}) z^2
\label{potential-taylor}
\end{equation} 
where $\bm{x}=(x,0,0)$ and 
\begin{eqnarray}
\label{potentialx}
U_\alpha(\bm{x}) &=& -\frac{\hbar \Gamma_\alpha^2}
{4 \pi I_\alpha^{\mathrm sat}} 
\left[
\frac{T_1^\alpha P_1}{w_1^2} \frac{1}{ 1+x^2/R_1^2}
\right. \nonumber \\ &&\qquad + \left.
\frac{T_2^\alpha P_2}{w_2^2} 
\exp \left(-2x^2/w_2^2\right)
\right].
\end{eqnarray}
The expression for $U_\alpha(\bm{x})$ explicitly shows that 
there is a soft attractive potential on the 
Rayleigh range scale $R_1$ and a sharp repulsion around the origin 
on the beam waist lengthscale $w_2$. 
The net effect of the beams is to establish a double well potential 
along the $x$-axis with minima at $\pm \bm{x}_m = (\pm x_m,0,0)$
having at the same time a strong quasi harmonic confinement in the $y-z$ plane.
The transverse angular frequencies 
$\omega_{\alpha y}(\bm{x}) = (\partial^2_yU_\alpha(\bm{x})/m_\alpha)^{1/2}$ 
and 
$\omega_{\alpha z}(\bm{x}) = (\partial^2_zU_\alpha(\bm{x})/m_\alpha)^{1/2}$ 
are one order of magnitude larger than the intra-well longitudinal 
angular frequency 
$\omega_{\alpha x}(\bm{x}_m) = 
(\partial^2_xU_\alpha(\bm{x}_m) /m_\alpha)^{1/2}$. 
For atomic gases with chemical potential $\mu_\alpha$ satisfying 
$\hbar \omega_{\alpha x} \ll \mu_\alpha \ll \hbar \omega_{\alpha y}, 
\hbar \omega_{\alpha z}$ the present configuration 
thus realizes a quasi-1D trapped gas analogously to that 
obtained in highly elongated magnetic traps \cite{Gorlitz}.

\begin{figure}
\psfrag{x}[bl][]{$x$ ($\mu$m)}
\psfrag{y}[r][]{$y$ ($\mu$m)}
\psfrag{z}[l][]{$z$ ($\mu$m)}
\psfrag{20}{20}
\psfrag{10}{10}
\psfrag{0}{0}
\psfrag{5}{5}
\psfrag{-10}{-10}
\psfrag{-20}{-20}
\psfrag{0.6}{0.6}
\psfrag{-0.6}{-0.6}
\includegraphics[width=0.8\columnwidth,clip]{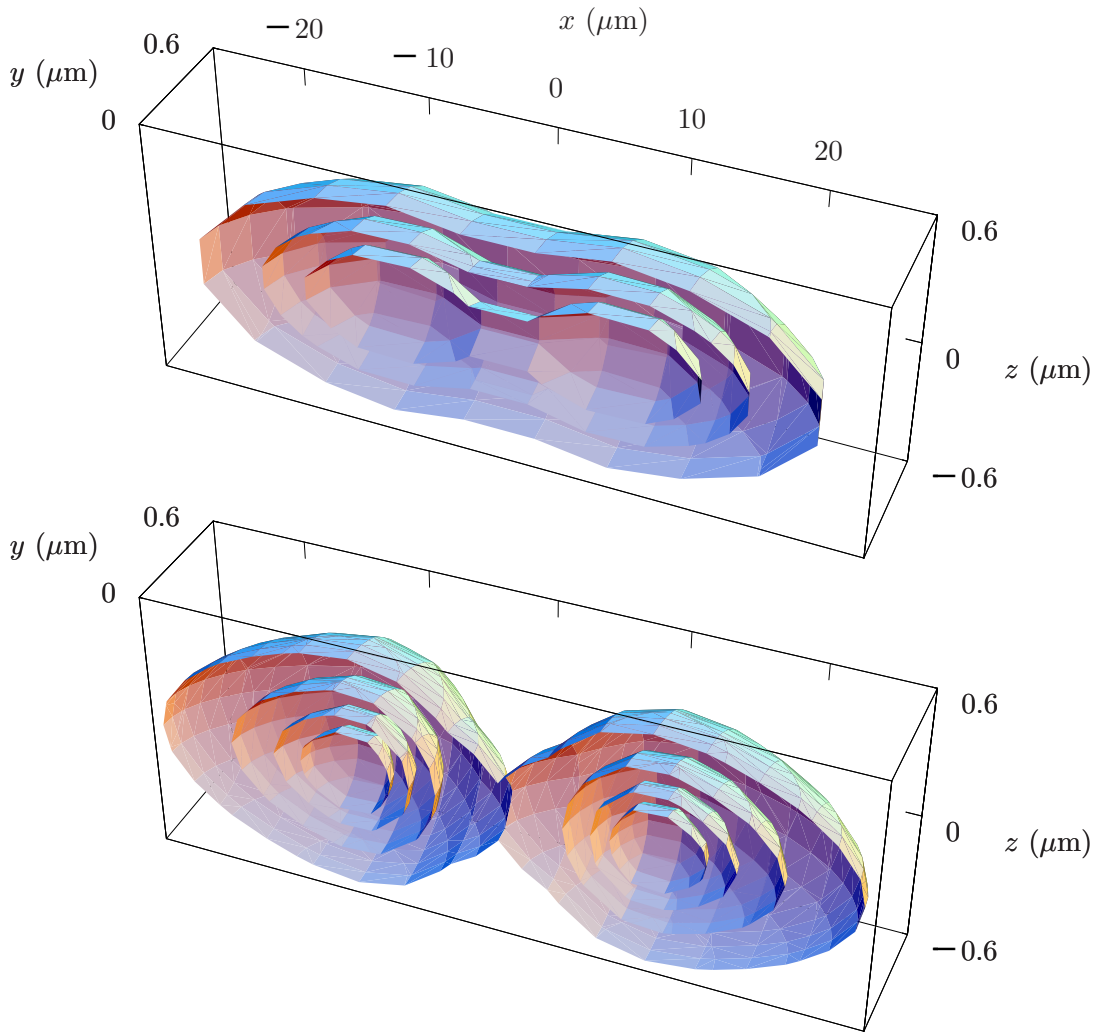}
\caption{
Bistable potential for optically trapped Fermi-Bose mixtures. 
The plots represent the equipotential surfaces above the minima 
of the potential $U_\alpha^\mathrm{min}$ by an amount 
$\Delta U_\alpha=$0.5, 2.5, 5, 10, 20 nK (from inner to outer shells, 
respectively) for fermionic $^6$Li (top) and bosonic $^{23}$Na (bottom).
We assume a laser power of $P_1=10$ mW at $\lambda_1$=1064 nm, 
$P_2/P_1=2.5 \times 10^{-3}$ at $\lambda_2$=532 nm, 
and waists $w_1=w_2=10~ \mu$m.
The atomic transition wavelengths are $\lambda_\mathrm{f}=671$ nm and 
$\lambda_\mathrm{b}=589$ nm.
}
\label{fig1}
\end{figure}

\begin{figure*}
\psfrag{x}[b][]{$x$ ($\mu$m)}
\psfrag{y}[b][]{$y$ ($\mu$m)}
\psfrag{U}[t][]{$U_\alpha(x,0,0)$ (nK)}
\psfrag{Na}[][]{$^{23}$Na}
\psfrag{Li}[][]{$^6$Li}
\psfrag{min}[l][]{0.5 nK}
\psfrag{max}[l][]{70 nK}
\psfrag{70}{70}
\psfrag{20}{20}
\psfrag{10}{10}
\psfrag{0}{0}
\psfrag{5}{5}
\psfrag{-10}{-10}
\psfrag{-20}{-20}
\psfrag{0.6}{0.6}
\psfrag{-0.6}{-0.6}
\psfrag{0.5}{0.5}
\psfrag{-0.5}{-0.5}
\psfrag{1}{1}
\psfrag{-1}{-1}
\includegraphics[width=1.0\columnwidth,clip]{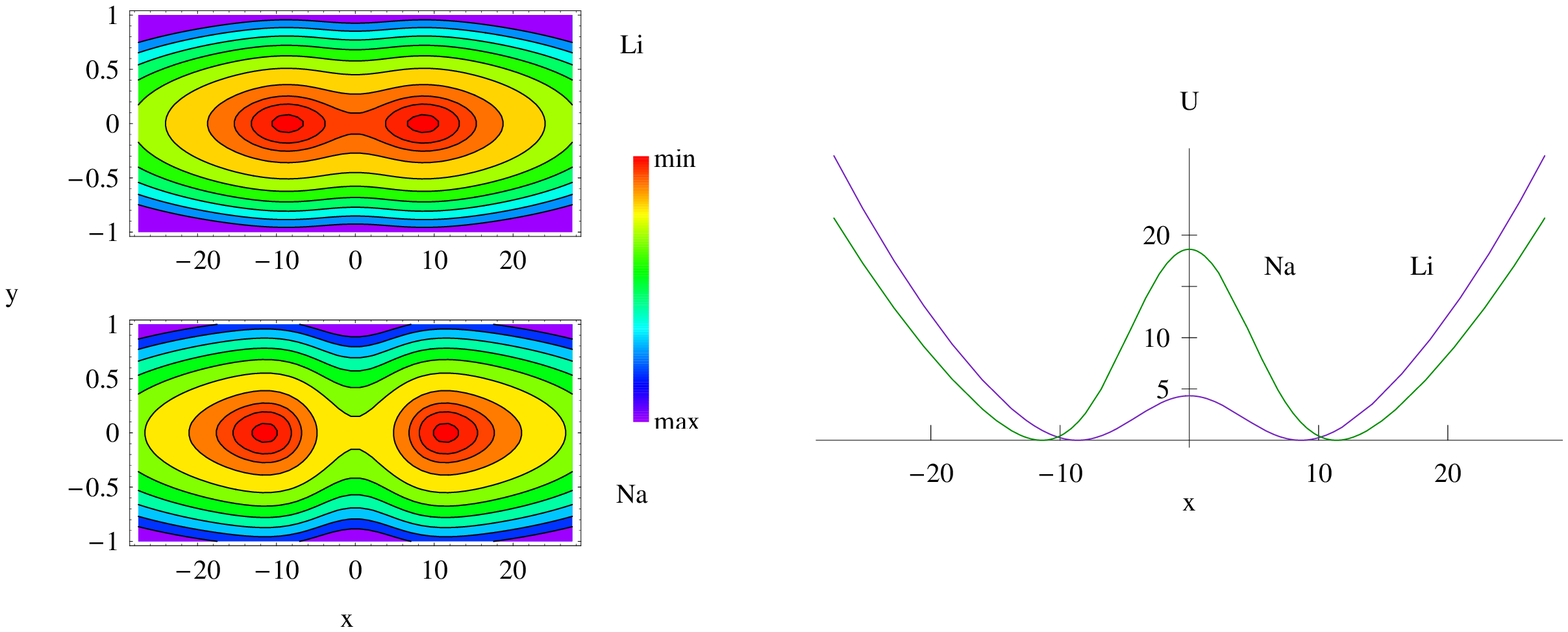}
\caption{Equipotential profiles in the $x-y$ plane of $U_\alpha(x,y,0)$ 
(left),  
for $\Delta U_\alpha=$0.5, 2.5, 5, 10, 20, 30, 40, 50, 60, 70 nK 
(from inner to outer shells, respectively),
and potential energy along the $x$ axis for the two species (right). 
The minima of the bistable potentials have been shifted to
zero for the sake of comparison, their values being 
$U_\mathrm{f}^\mathrm{min}=-3.87~ \mu$K and 
$U_\mathrm{b}^\mathrm{min}=-3.31~ \mu$K.
All trap parameters as in Fig. \ref{fig1}.}
\label{fig2}
\end{figure*}

In Fig. \ref{fig1} we show the potential energy for the Fermi and 
Bose components in the case of the $^6$Li-$^{23}$Na mixture, 
already brought to degenerate regime \cite{Hadzibabic}, by assuming 
a Nd:YAG laser emitting at $\lambda_1=1064$~nm as red-detuned source, 
and its second harmonic as blue-detuned source.  
Details of the potential energy profiles are shown for the two
species in Fig. \ref{fig2}. 
It is evident that the bosonic species experiences a 
double well potential with larger distance and higher potential barrier
between the minima with respect to the fermionic one.

Some comments are in order.
The Fermi gas is always more deeply and strongly confined 
than the Bose gas, which is favorable for different reasons. 
Firstly, this allows for a continuous evaporative cooling of the Bose 
gas marginally affecting the Fermi gas. 
Secondly, the ratio between the average trapping frequency for 
fermions and bosons maintains a value greater than unity, with 
a negligible decrease with respect to the case of coaxial beams.  
This allows for efficient sympathetic cooling of a degenerate Fermi 
gas and a less degenerate Bose gas, through matching of the 
specific heats. Indeed the Bose gas, less degenerate, maintains 
its large classical specific heat at lower temperatures with 
respect to the more degenerate Fermi gas \cite{Presilla}. 
Thirdly, this reduces the peak density of the Bose gas at 
the center of the trap making less relevant both the boson-boson interaction, 
and most importantly the fermion-boson interaction. 

Finally, the fact that bosons experience a higher potential 
barrier and a larger separation between the minima, 
strongly differentiates the dynamics of the two species.  
As explained in the following, the latter feature allows for 
two unambiguous signatures of the macroscopic coherence associated 
to a possible superfluid phase of the Fermi gas, 
based on tunneling oscillations and interference, respectively.

\section{Tunneling phenomena} 
Macroscopic tunneling phenomena have been 
successfully explored in a Fermi liquid, namely $^3$He \cite{Avenel}, and 
it is therefore natural to explore their counterpart in dilute Fermi gases. 
In our configuration, by using sufficiently low barriers the tunneling
probability for fermions may become large enough to make the 
detection of the corresponding interwell oscillation feasible. 
Here there is a rich scenario due to the possibility of single-particle 
tunneling for both fermions in a degenerate but normal state and bosons in a
thermal state (either quantum or thermally activated), and macroscopic 
quantum tunneling of fermions in a BCS state and of bosons in a BEC state. 
An oscillating tunneling current between the two wells of the trap is obtained 
by breaking the symmetry along the $x$ axis by means of a tilting potential, 
$V(x)=b ~x$, suddenly added to the trapping potential $U_\alpha(x,y,z)$, 
e.g. using Zeeman shifts generated by a quadrupole magnetic field with 
symmetry axis along the $x$ direction. 

The evaluation of the macroscopic tunneling current is particularly 
simple within the quasi one-dimensional approximation 
(\ref{potential-taylor}) of the potential.
Suppose that the total number $N_\alpha$ of trapped 
(bosons or bosonized fermions) atoms avaliable for macroscopic tunneling
is in the ground state of the potential (\ref{potentialx}) and let
$\psi_\alpha(x)$ be the unit-normalized associated wavefunction.
If a tilting potential $V(x)=b ~x$ with small bias parameter $b$ 
is added at time $t=0$, 
the evolution of the system wavefunction can be obtained as 
\begin{equation}
\psi_\alpha(x,t)=c_1^\alpha e^{iE_1^\alpha t/\hbar} \phi_1^\alpha(x)
+c_2^\alpha e^{iE_2^\alpha t/\hbar} \phi_2^\alpha(x),
\label{psit}
\end{equation}
where $\phi_1^\alpha(x)$ and $\phi_2^\alpha(x)$ are the first two 
eigenfunctions,
with eigenvalues $E_1^\alpha$ and $E_2^\alpha$, respectively,
of the tilted potential $U_\alpha(\bm{x})+V(x)$
and the coefficients $c_1^\alpha$ and $c_2^\alpha$ are given by
\begin{equation} 
c_i^\alpha = \int_{-\infty}^{+\infty} 
\phi_i^\alpha(x) \psi_\alpha(x) dx, \qquad i=1,2 . 
\end{equation}
In the absence of decoherence phenomena the evolution of the macroscopic
wavefunction (\ref{psit}) gives rise to condensate fractions
in the left and right wells, $N_\alpha^L(t)$ and $N_\alpha^R(t)$ 
oscillating in time with $N_\alpha^L(t) + N_\alpha^R(t) = N_\alpha$.
The corresponding current is easily evaluated as
\begin{equation} 
\frac{dN_\alpha^L}{dt} =N_\alpha 
\frac{d}{dt}\int_{-\infty}^0|\psi_\alpha(x,t)|^2 dx=
A_\alpha \sin \frac{\Delta E_\alpha t}{\hbar},
\end{equation}
with $A_\alpha=2 c_1^\alpha c_2^\alpha c_{12}^\alpha N_\alpha  
\Delta E_\alpha/\hbar$ the amplitude of the tunneling current, 
depending upon the energy splitting $\Delta
E_\alpha=E_2^\alpha-E_1^\alpha$, and 
\begin{equation}
c_{12}^\alpha=\int_{-\infty}^{0} \phi_1^\alpha(x) 
\phi_2^\alpha(x)dx
\end{equation}
is the overlap integral. 

To evaluate the macroscopic tunneling current for the 
realistic trap potential (\ref{potential-taylor}), the eigenvalues 
$E_1^\alpha$ and $E_2^\alpha$ and the corresponding eigenfunctions
$\phi_1^\alpha$ and $\phi_2^\alpha$ must be found numerically.
Due to the strongly differentiated tunneling dynamics for bosons 
and bosonized fermions, we have the hard numerical problem of finding
exceedingly small energy splittings.
The selective relaxation algorithm proposed in \cite{Pretambo}
allows to solve the problem at least in the parameter region of 
physical interest. 

\begin{figure}
\psfrag{x}[t][]{$b$ (nK/$\mu$m)}
\psfrag{y}[b][]{$A_\alpha/N_\alpha$, $\Delta E_\alpha/ \hbar$ (s$^{-1}$)}
\psfrag{Li}{$^6$Li-$^6$Li}
\psfrag{Na}{$^{23}$Na}
\includegraphics[width=0.85\columnwidth,clip]{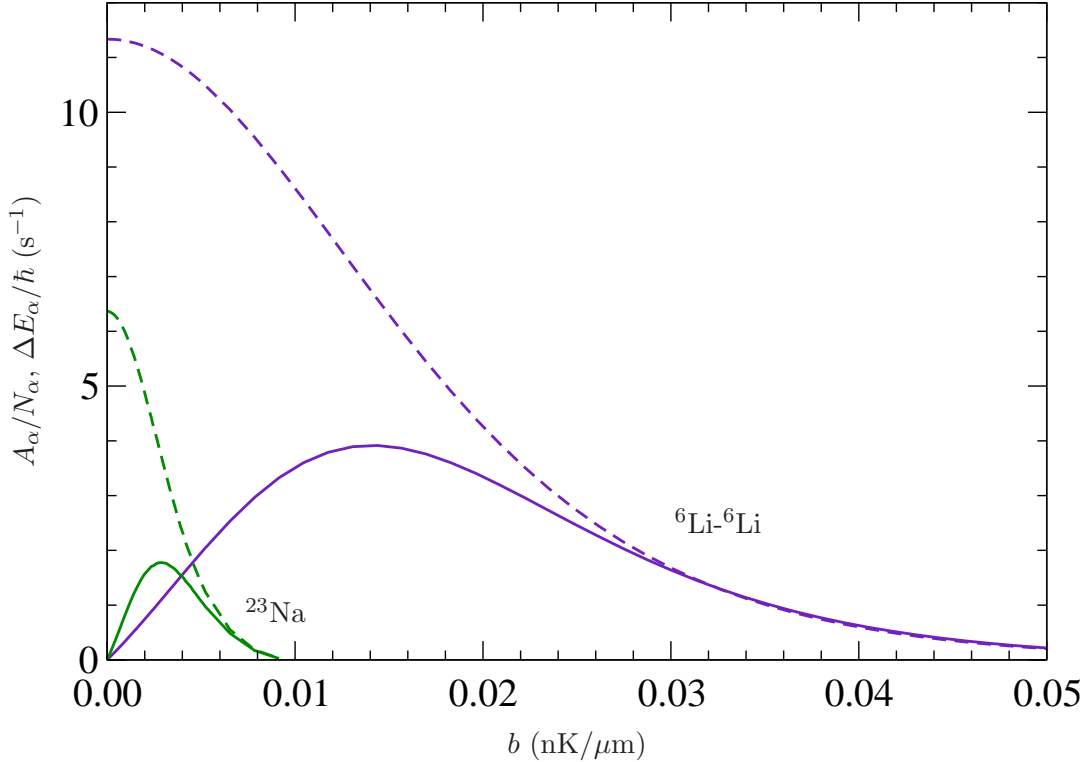}
\caption{
Macroscopic coherence of the Fermi gas through tunneling experiments.
Dependence of tunneling current amplitude per unit of available atoms 
$A_\alpha/N_\alpha$ (solid line) and tunneling angular frequency 
$\Delta E_\alpha/\hbar$ (dashed line) for $^{23}$Na and $^6$Li-$^6$Li 
Cooper pairs versus bias strength $b$. 
For the sake of comparison, amplitude and frequency of bosons have 
been multiplied by $10^5$. The trap parameters are as in Fig. \ref{fig1}.}
\label{fig3}
\end{figure}

In Fig. \ref{fig3} we report the amplitude and the angular frequency of the 
currents for macroscopic tunneling of $^6$Li-$^6$Li Cooper pairs and 
of the Bose condensed component of $^{23}$Na versus the tilting parameter $b$. 
For both species there is an optimum value of the bias maximizing the 
observability of tunneling oscillations, with a maximum value
$A_\alpha \simeq 4$ s${}^{-1}$ obtained for a bias value of the
tilting potential $b \simeq 14$ pK/$\mu$m. 
Tunneling current and frequency for the Bose gas are smaller by five orders 
of magnitude with respect to the analogous quantities of the Cooper pair gas. 
This results from the exponential sensitivity of quantum tunneling to the 
different potentials experienced by the two species and to their different 
masses.

Provided that the number of Cooper pairs is large enough, the
modulation of the number of atoms in the two wells can be evidenced 
by using non-destructive phase-contrast imaging.
The low intensity of the blue-detuned beam required to create a
tunneling regime in the presence of a magnetic trap with limited
confinement strength prevented observation of tunneling phenomena in the 
experiment described in \cite{Andrews} (see also \cite{Shin} for 
a recently achieved bistable configuration). 
In the situation proposed here this issue is circumvented due to the 
possibility to change both the intensities of the beams while maintaining 
their ratio constant, at least to the extent that heating 
from residual Rayleigh scattering does not play a significant role. 
Current techniques allow for relative stabilization at the 1\% level
or below (for a general discussion of laser stabilization techniques
see \cite{Ralph}), especially for frequency-doubled beams as in the 
proposed configuration.
 
An important requirement for this proposed test is to maintain the decoherence 
rate low enough to minimize damping of the coherent oscillations which, according 
to Fig. 3, are expected to occur with periods of order $100$ ms or longer. 
In the cooling strategy outlined in \cite{Presilla} there is not
a strict need to use enhancement of the elastic scattering length through Feshbach
resonances \cite{Holland,Timmermans} to reach a deep Fermi degenerate
regime. This could circumvent the issue of decoherence sources due to
enhancement of density, like those discussed in \cite{Jack} for
three-body collisions. Thus decoherence is mainly expected from the
presence of Rayleigh heating or more technical sources like intensity
and beam pointing fluctuations of the laser beams. Their effect is a 
temperature increase with a consequent increase of the thermal component 
\cite{Graham} and the single-particle tunneling current of bosons and fermions.
It should be noted however that in an optical dipole trap, due to the
smaller trapping volume and lower trap depth, we expect a suppressed
thermal fraction. This is a further advantage in using an optical 
trap with respect to a magnetic trap for coherence experiments. 
Continuous evaporative cooling of the Bose species should also
mitigate its effect. 
One can then envisage a cooling strategy where coherent oscillations 
persist for a much longer time at time-dependent amplitude and
frequency, therefore originating a chirped signal.

\section{Interference phenomena} 
In a landmark experiment, 
the Ketterle group evidenced the macroscopic coherence of a pair of 
Bose condensates by looking at the interference fringes resulting after 
their release from a bistable potential \cite{Andrews}. 
The latter was obtained by the combination of a harmonic potential 
created by  a magnetic trap, and a blue-detuned beam focused on the 
magnetic potential minimum with propagation orthogonal to the weaker 
confining axis.  
The average distance between the two condensates was controlled by
changing the power of the blue-detuned beam. 

\begin{figure}
\psfrag{x}[t][]{$P_2/P_1$}
\psfrag{y}[b][]{$\ell_\alpha$ ($\mu$m)}
\psfrag{Li}{$^6$Li-$^6$Li}
\psfrag{Na}{$^{23}$Na}
\includegraphics[width=0.85\columnwidth,clip]{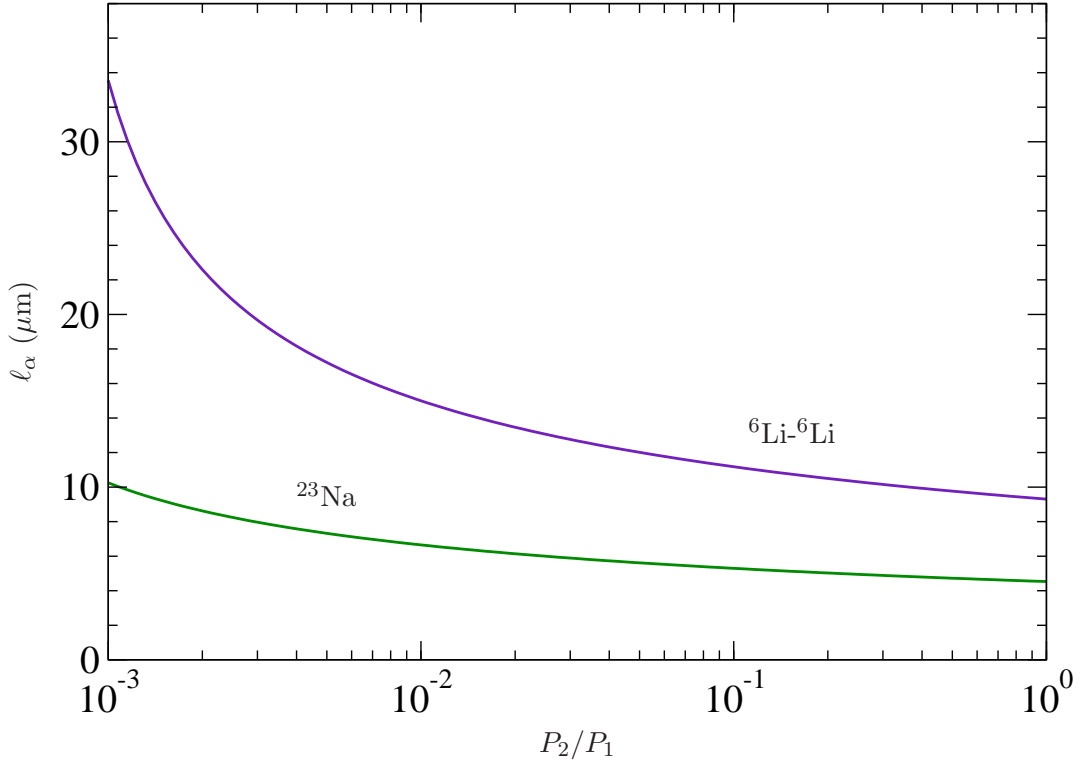}
\caption{
Macroscopic coherence of the Fermi gas through interference experiments.
Dependence of fringe spacing $\ell_\alpha$ for 
$^{23}$Na bosons and $^6$Li-$^6$Li Cooper pairs versus 
$P_2/P_1$. We assume a time of flight $t=10$ ms.}
\label{fig4}
\end{figure}

A similar experiment can be repeated with an ultracold Fermi-Bose
mixture in a bistable optical dipole trap. 
Above the temperature for the onset of a BCS-like phase 
transition of the Fermi gas one expects only interference fringes 
arising from the condensed fraction of the Bose gas. 
Below $T_\mathrm{BCS}$ we do expect also the emergence of an interference 
pattern coming from the bosonized fermions, a small fraction of the 
total number of Fermi atoms. The distance between the peaks of maximum 
signal in the interference pattern is given, for a free expansion, by 
$\ell_\alpha=2\pi \hbar t/m_\alpha d_\alpha$. Here $t$ is the time of flight, 
$d_\alpha$ the initial distance between the centers of mass of the
two clouds before the release from the trap, and 
$m_\alpha$ either the mass of the Bose atoms or twice the mass of each 
fermion in the case of the BCS-bosonized component.
Since the Fermi species in our example has both a significantly 
smaller mass and separation $d_\alpha$ than the Bose species we do expect
an easy discrimination of the interference pattern attributable to the 
former species.  This is confirmed by looking at the distance 
between the interference peaks for $^6$Li and $^{23}$Na 
versus the $P_2/P_1$ power ratio, as depicted in Fig. \ref{fig4}.  
One can take advantage of this dependence of the interference patterns 
to discriminate any effect of the Bose component. 
The spacing of the interference fringes expected from the macroscopic 
wavefunction associated to the bosonized Fermi component maintains a 
value $\sim 2$ times larger than the corresponding one for the Bose species.
Selective optical pumping tomography on the Fermi species cycling transition  
as in \cite{Andrews} can allow to enhance the corresponding interference 
signal. 
Considering the very dilute nature of the bosonized Fermi gas, 
as a consequence of the less stiff confinement of the Bose species, we do
not expect a significant decrease of the fringe visibility due to the 
mean field effects as instead already observed for bosons
\cite{Rohrl}. Due to the lower trapping frequencies when using a 
bichromatic optical trap as discussed in \cite{Onofrio}, the mean-field 
effects due to the interaction between the Fermi and the Bose gas are 
also negligible. 

\begin{figure}
\includegraphics[width=0.95\columnwidth,clip]{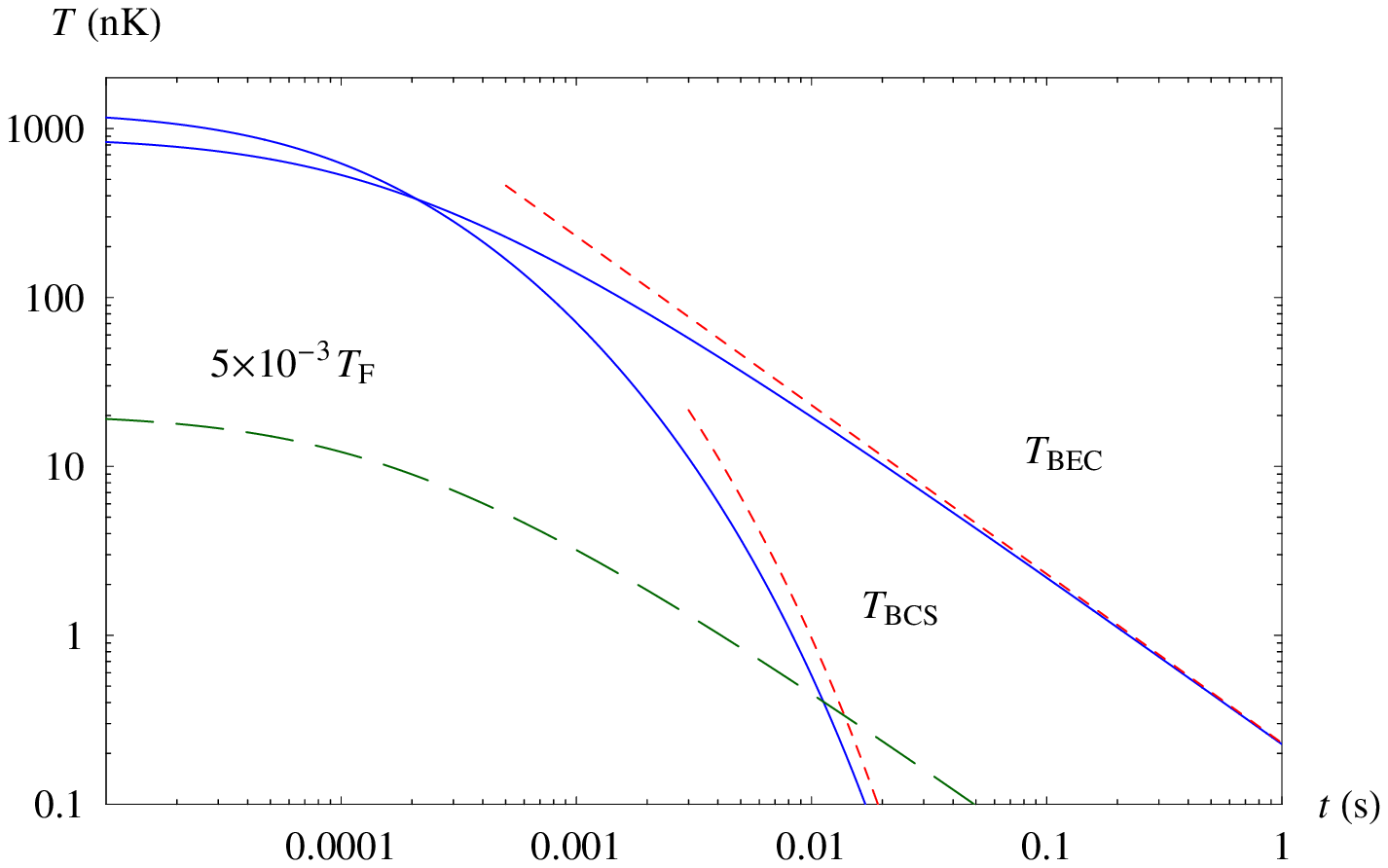}
\caption{
Temperature of the cloud (dashed line), and critical temperatures 
$T_\mathrm{BCS}$, $T_\mathrm{BEC}$ (continuous lines) 
as a function of time $t$ in the case of a freely expanding $^6$Li 
cloud with $N_\mathrm{F}=3 \times 10^6$ atoms. 
The initial density is $n_0= 3.5 \times 10^{13}$ cm$^{-3}$ and 
we use an elastic scattering length for fermions $a=-230$ nm \cite{Stoof}. 
The dashed lines close to the critical temperature continuous lines 
are the asymptotic values for $T_\mathrm{BCS}$, 
$T_\mathrm{BEC}$ obtained by substituting $n(t)$ 
with $n_\infty(t)$ into (\ref{TBCS}) and (\ref{TBEC}).}
\label{fig5}
\end{figure}

The interference experiment discussed above could also allow to distinguish 
if the Fermi gas is in a BCS or BEC regime. 
For BCS macroscopic states we do expect a loss of fringe contrast 
at expansion times much smaller than those reachable in the case of molecular 
BEC states. This behaviour is quantitatively shown by assuming an
adiabatic expansion of the gas after the release from the trap, and a rough 
estimate of the critical temperatures, $T_\mathrm{BCS}$ and 
$T_\mathrm{BEC}$ as follows.
Suppose that at time $t=0$, when the trap potential is turned off, 
the gas made of $N_\mathrm{F}$ fermions of mass $m_\mathrm{F}$ 
has density $n_0$. 
Let $R_0$ be the radius of the equivalent sphere containing the gas at
$t=0$ defined by $\frac{4}{3} \pi R_0^3 n_0 = N_\mathrm{F}$. 
At later times the cloud expands and, assuming an ideal behavior,
the radius of the equivalent sphere is given by the law 
$R(t)=R_0+ v_\mathrm{F}t$. 
As a consequence, the density of the gas $n(t)$ decreases from its initial
value $n_0$ and, in turn, the Fermi velocity 
$v_\mathrm{F}= \hbar (3 \pi^2 n)^{1/3}/m_\mathrm{F}$ also decreases.
The density of the gas is therefore determined by the following 
self-consistent equation
\begin{equation}
n(t) = \frac{N_\mathrm{F}}
{\frac43 \pi \left[ R_0+\frac{\hbar t}{m_\mathrm{F}} 
(3 \pi^2 n(t))^{\frac13} \right]^3},
\label{nself}
\end{equation}
which, for a generic value of $t$, must be solved numerically.
For $t \gg t_0 = m_\mathrm{F} R_0 / \hbar (3\pi^2 n_0)^{1/3}$, 
the initial radius $R_0$ can be neglected in the denominator of (\ref{nself}) 
and we obtain the asymptotic time-varying density 
\begin{equation}
n_\infty(t) = 
\sqrt{ \frac{n_0}{3\pi^2} \left(\frac{m_\mathrm{F}R_0}{\hbar t}\right)^3 }.
\label{ninf}
\end{equation}
Once the density $n(t)$ is known, we have an explicit estimate 
of the critical temperature for BCS transition \cite{Stoof} 
as a function of time
\begin{equation}
T_\mathrm{BCS}(t) = \frac{5}{3e} 
\frac{\hbar^2(3 \pi^2 n(t))^{\frac23}}{2 m_\mathrm{F} k_\mathrm{B}}
\exp \left[ -\frac{\pi}{ 2|a| (3\pi n(t))^{\frac13} } \right],
\label{TBCS}
\end{equation}
where $a$ is the elastic scattering length of the fermionic species.
The critical temperature $T_\mathrm{BEC}$ for fermions condensed via
molecular states can be estimated from the critical temperature of an 
ideal gas of bosons \cite{Dalfovo}
with mass $2m_\mathrm{F}$ and density $n(t)/2$
\begin{equation}
T_\mathrm{BEC}(t) = 
\frac{\pi \hbar^2}{m_\mathrm{F} k_\mathrm{B}} 
\left( \frac{n(t)}{2\zeta(3/2)} \right)^{2/3},
\label{TBEC}
\end{equation}
where $\zeta$ is the Riemann zeta function.

In Fig. \ref{fig5} we show the behavior of the critical temperatures 
$T_\mathrm{BCS}$ and $T_\mathrm{BEC}$, and of the temperature of the 
cloud assuming an adiabatic expansion \cite{Menotti}, 
versus time in the case of $^6$Li.
While $T_\mathrm{BEC}$ has only a $n(t)^{2/3}$ dependence,
and therefore decreases for large times as $t^{-1}$,
$T_\mathrm{BCS}$ has a further exponential suppression factor
and for large $t$ decreases as $t^{-1}\exp \left(-\sqrt{t}\right)$.
By assuming an initial temperature $T=5 \times 10^{-3} T_\mathrm{F}$, 
the disappearance of the fringes at times $t \sim 10$ ms when 
$T/T_\mathrm{BCS}>1$  would imply a BCS regime for the degenerate Fermi gas.
On the other hand, the observation of interference effects should be
possible up to longer times in the case of fermions coupled 
through molecular BEC, since the scaling of the temperature of the
cloud and $T_\mathrm{BEC}$ are similar. 

\section{Conclusions} 
We have discussed a quasi one-dimensional bistable configuration for 
optically trapped atoms. Two signatures have been discussed 
for evidencing macroscopic quantum coherence of a paired component of
a Fermi gas, regardless of the BEC or BCS-like regime for the Fermi
gas. Our proposal allows to identify a superfluid component both in a strongly-coupled 
regime based upon enhancement of scattering length (molecular BEC regime) or 
in a BCS-like regime obtained by just cooling the sample at very low temperatures 
as suggested in \cite{Presilla}, with efficient heat capacity
matching between the Fermi and the Bose species, 
without necessarily exploiting Feshbach resonances to obtain
large critical temperatures for BCS pairing, although their 
use is certainly possible in our framework. 
The use of a Fermi-Bose mixture seems preferable since 
analogous quantum coherence experiments involving mixtures 
of two Zeeman levels of fermions are less easy to perform.
Indeed, two distinguishable Fermi states with same mass  
will give rise to two independent interference patterns 
or tunneling currents just differing by the initial random phase, 
unless a locking mechanism is used.  Also, dual evaporative 
cooling substantially decreases the number of atoms potentially 
available for Cooper pairing, then sensibly diminishing the expected signal.  
Moreover, the presence of the Bose gas until the last stage of the 
cooling is useful to quantitatively assess the temperature of the 
Fermi gas, and allows to study a variety of situations for which 
changes to the BCS-coupled Fermi pairs are expected when bosons 
mediates their interactions \cite{Heiselberg}. 
 
There have been many refined and ingenious proposals for the observation 
of the superfluid phase of an ultracold Fermi gas, ranging
from the study of collective modes \cite{Baranov}, moment of inertia
\cite{Farine}, density profile of the Cooper-paired component \cite{Chiofalo}, 
light scattering \cite{Zhang,Torma,Roustekoski,Weig,Wong,Search}, 
free expansion  \cite{Menotti}, Bloch oscillations 
\cite{Rodriguez}, internal Josephson effect \cite{Paraoanu}, 
and Raman photoassociation in Bose-Fermi mixtures \cite{Mackie}. 
All these proposals, including ours, will have to face the small number of fermions 
available in the superfluid state, and the subsequent small 
signal-to-noise ratio for any conceivable signature of the phase transition. 
In view of these experimental difficulties it is crucial to seek for 
redundancy of signatures with diverse techniques, hopefully all 
converging in individuating a common superfluid phase diagram.   

\begin{acknowledgments}
We thank M. L. Chiofalo for useful discussions, and L. Viola 
for a critical reading of the manuscript. This work was supported 
in part by Cofinanziamento MIUR protocollo 2002027798$\_$001.
\end{acknowledgments}

\end{document}